# Zimin patterns in genomes

Nikol Chantzi[1], Ioannis Mouratidis[1], Ilias Georgakopoulos-Soares[1,*]

[1] Institute for Personalized Medicine, Department of Biochemistry and Molecular Biology, The Pennsylvania State University College of Medicine, Hershey, PA, USA.
*correspondence to: izg5139@psu.edu

**Abstract**

Zimin words are words that have the same prefix and suffix. They are unavoidable patterns, with all sufficiently large strings encompassing them. Here, we examine for the first time the presence of k-mers not containing any Zimin patterns, defined hereafter as Zimin avoidmers, in the human genome. We report that in the reference human genome all k-mers above 104 base-pairs contain Zimin words. We find that Zimin avoidmers are most enriched in coding and Human Satellite 1 regions in the human genome. Zimin avoidmers display a depletion of germline insertions and deletions relative to surrounding genomic areas. We also apply our methodology in the genomes of another eight model organisms from all three domains of life, finding large differences in their Zimin avoidmer frequencies and their genomic localization preferences. We observe that Zimin avoidmers exhibit the highest genomic density in prokaryotic organisms, with *E. coli* showing particularly high levels, while the lowest density is found in eukaryotic organisms, with *D. rerio* having the lowest. Among the studied genomes the longest k-mer length at which Zimin avoidmers are observed is that of S*. cerevisiae* at k-mer length of 115 base-pairs. We conclude that Zimin avoidmers display inhomogeneous distributions in organismal genomes, have intricate properties including lower insertion and deletion rates, and disappear faster than the theoretical expected k-mer length, across the organismal genomes studied.

## Introduction

Examining patterns that are absent from the genome space of an organism can reveal valuable biological knowledge including insights in sequence composition and evolution, genomic constraints and pathogenicity as well as hints regarding the underlying biological mechanisms. Previous work has showcased the shortest k-mer sequences that are absent from a genome, termed nullomers (Hampikian and Andersen, n.d.). Another related concept is minimal absent words, which are nullomers, but removing either their leftmost or rightmost character generates a k-mer that is present in the genome (Pinho et al. 2009; Garcia and Pinho 2011; Almirantis et al. 2017). Nullomers and minimal absent words have since been used in a plethora of applications, including cancer detection, pathogen surveillance, phylogenetic tree construction, drug development and in forensic science among others (Tsiatsianis et al. 2024; Montgomery et al. 2024; Alileche and Hampikian 2017; Ali et al. 2024; Goswami et al. 2013; Mouratidis et al. 2023; Koulouras and Frith 2021; Georgakopoulos-Soares et al. 2021).

A number of research papers have successfully applied theoretical mathematical concepts in genomics, such as the golden ratio, fractals and the Fibonnaci numbers among others (Persaud and O'Leary 2015; Yamagishi and Shimabukuro 2008; Stanley et al. 1999). Of high interest is the field of combinatorics as its theory and concepts can be applied in genomics in k-mer based investigations. A sesquipower or Zimin word is a sequence of characters over an alphabet with identical prefix and suffix, a concept that was proposed by A. I. Zimin over four decades ago (Zimin 1984). Zimin words are unavoidable patterns, and all sufficiently long strings must contain Zimin words. The field of Zimin words has grown considerably since its inception and contains a sizable research literature (Choffrut and Karhumäki 1997; Lothaire 2014; GöllerStefan 2019; Cooper and Rorabaugh 2014). Nevertheless, Zimin words have never been examined in bioinformatics using the DNA alphabet and their identification and characterization in organismal genomes has not been performed to date. Provided that over half of the human genome contains repetitive sequences and patterns (de Koning et al. 2011), the characterization of Zimin words and Zimin avoiding k-mer sequences in the human genome, and in the genome of other organisms, can provide insights into genome organization, structure and grammar.

Here, we identify Zimin words, from which we derive Zimin avoiding sequences, across nine organismal genomes. In this context, we will refer to Zimin avoiding sequences as $Z_3$ *avoidmers*, implying exclusive $Z_3$ avoidance, i.e. the k-mers that avoid, or, equivalently, do not contain any subsequence which captures the *abacaba* pattern on the four letter nucleotide alphabet *{a, g, c, t}* (see Definitions). We report that the upper k-mer limit after which, $Z_3$ Zimin avoiding sequences cannot be detected in the reference human genome is 105 base-pairs (bps). We also show that Zimin avoidmers are less frequent than expected. Zimin avoiding sequences are enriched in coding regions and classic human satellite hsat1B regions in the human genome and are depleted for germline insertion and deletion (indel) variants relative to their surrounding regions. This study provides the first biological application of Zimin words. Future work is required to further investigate their properties, potential roles in organismal genomes and their usage in the development of novel tools and applications.

## Definitions

An *alphabet L* is a collection of symbols often referred to as *letters*. Let $L = \{a_1, \ldots, a_m\}$ an alphabet of *m* letters. A *word w* is an ordered sequence of letters drawn from the alphabet $L$. We denote by $L^*$ the set of all words over $L$. For any word $w \in L^*$ the length of $w$ is denoted by $|w|$. Additionally, we denote by $\varepsilon$ the *empty word*. By definition, $|\varepsilon| = 0$.

In genomics, the nucleotide alphabet is defined naturally as the set of all four nucleotides, i.e. $L = \{a, g, c, t\}$. Small finite words of the nucleotide alphabet of length k are often referred to as *k-mers*. For instance, $w = aagtaag$ is a 7-mer.

A *homomorphism* $\varphi: L_1^* \to L_2^*$ is a function between two non-empty sets $L_1^*$ and $L_2^*$ such that:
$$\varphi(xy) = \varphi(x)\varphi(y),$$
for any words $x, y \in L_1^*$. A homomorphism $\varphi$ is said to be *non-erasing*, if $\varphi(x) \neq \varepsilon$ for any $x \in L_1^*$.

Building on the foundation from (Carayol and Göller 2019), let $V = \{v_1, v_2, \ldots, v_n\}$ a finite set of variables. A *pattern P* is a finite word over $V$. We will say that a word w encounters pattern P, if there is a subsequence z of w and a non-erasing homomorphism φ such that $\varphi(P) = z$.

For example, consider a set of pattern variables $V = \{x, y\}$ and the nucleotide alphabet $L = \{a, g, c, t\}$. The sequence $w = aagtaagaag$ encounters the pattern Q=xyxx, since there exists a non-erasing homomorphism $\varphi: V^* \to L^*$ defined by $x \to aag$ and $y \to t$, which embeds pattern *xyxx* into w, since
$$\varphi(Q) = \varphi(xyxx) = \varphi(x)\varphi(y)\varphi(x)\varphi(x) = aagtaagaag = w.$$

## Definition of unavoidability

A pattern $P$ is q-*unavoidable* if all but finitely many finite words over an alphabet of size q, encounter P. A pattern $P$ is unavoidable if it is *q-unavoidable* for all $q \geq 1$.

For instance, the pattern $P = xx$ is unavoidable in the binary alphabet $q = 2$, since all binary words with at least 4 letters inevitably encounter it.

## Definition of Zimin words in DNA

Since there are four nucleotides in DNA, it is natural to shift our attention primarily to alphabets with four letters and, in particular, the nucleotide alphabet:
$$L = \{a, g, c, t\}.$$
Additionally, we will limit our attention to patterns that are exclusively 4-unavoidable.

For all $n \geq 1$, the $n$-th *Zimin pattern*, is defined recursively, as follows:
$$Z_0 = \varepsilon$$
$$Z_n = Z_{n-1} a_n Z_{n-1}, n \geq 1,$$

where $a_n$ denotes a distinct pattern variable. For instance, the first four Zimin patterns are defined as follows: $Z_1 = a$, $Z_2 = aba$, $Z_3 = abacaba$ and $Z_4 = abacabadabacaba$.

**Theorem** (Zimin 1984; Bean, Ehrenfeucht, and McNulty 1979)
A pattern $P$ containing $n$ different variables is unavoidable if, and only if, $Z_n$ encounters $P$.

The seminal theorem by Zimin (1984) has a corollary that over the nucleotide alphabet, any pattern of length n is unavoidable, i.e. any subsequence of sufficiently long k-mer will contain it, if and only if that pattern encounters $Z_n$. Structurally, this means that these patterns are bound to (re)occur within sufficiently long sequences. This triggered a whole new area of research to estimate asymptotically the upper and lower limits of the length of such sequences. Naturally, one would expect due to the repetitiveness of the human genome, such patterns to be rather scarce when compared to genomes that are less repetitive. Another important remark would be that:

In particular, a pattern $P$ containing three different variables *a,b,c* is unavoidable if, and only if, $Z_3 = abacaba$ encounters $P$. This naturally gives rise to the following definition of an avoidmer.

**Definition (Avoidmer)**
A k-mer w is called $Z_n$**-avoidmer** or **n-avoidmer** if it avoids Zimin $Z_n$ pattern. In particular, we will refer to a k-mer as **Zimin avoidmer**, or simply *avoidmer* to imply $Z_3$ avoidance.

A natural question to examine, is that given an alphabet *L* of size $|L| = q$, what is the smallest natural number $f(n, q)$ such that all the words of length $f(n, q)$ encounter the Zimin word $Z_n$. More formally, we define $f(n, q)$ as follows:

$$f(n, q) = min\{k \geq 1: w \text{ encounters } Z_n \text{ for any } w \in L \text{ such that } |w| \geq k\}.$$

Again, for our particular case, we will shift our attention to the nucleotide alphabet $q = 4$, and in particular for Zimin patterns $n \leq 3$.

We will list below some known results and properties and asymptotic upper bounds for $f(n,q)$:

- $f(1, q) = 1$
- $f(2, q) = 2q + 1$
- $f(3, q) \leq \sqrt{e}2^q(q + 1)! + 2q + 1$ ("Searching for Zimin Patterns" 2015)

In the case of the nucleotide alphabet $q = 4$, we have the following asymptotic bounds,
- $f(1,4) = 1$
- $f(2,4) = 9$
- $f(3,4) \leq 3,174$.

We denote by $\tilde{f}(3)$ the smallest *observed* natural number such that all k-mers that emerge in the human genome of length at least $\tilde{f}(3)$ encounter the Zimin word $Z_3$. By definition,
$$\tilde{f}(3) \leq f(3,4) \leq 3,174.$$

In fact, $\tilde{f}(3)$, in the human genome, is much lower than this upper bound, and unavoidance can be achieved much earlier at 105 bp. In addition, we note that, Zimin pattern *avoidance*, is a symmetric property, in the sense that, a kmer $w$ avoids $Z_n$, if and only if it's reverse complement *r(w)* avoids $Z_n$.

## Results

**Identification of Zimin avoidmers in the human genome**

It is established theoretically (Zimin 1984) that any sufficiently large word will entail $Z_3$ patterns in it. Thus, we examined the positional preferences of Zimin avoidmers across the human reference genome. We investigated the Zimin avoidmer distribution using the Telomere-to-Telomere (T2T) complete human genome (Nurk et al. 2022) across all chromosomes for $Z_3$ avoidmers, for k-mer lengths above 50bps. We find that the number of $Z_3$ avoidmers declines precipitously with k-mer length (**Figure 1a**). For instance, we report 4,651,253 $Z_3$ avoidmers of at least 50 bps length, while at 60 bps and 70 bps we report 675,325 and 70,546 $Z_3$ avoidmers, respectively.

We were interested to study how the distribution of the $Z_2$ and $Z_3$ avoidmer occurrences in the human genome compare to the expected distribution. For a given k-mer length, there are a total of $4^k$ possible k-mers. For the first 14 bp k-mer lengths, we exhaustively generated the subset of avoidmers and calculated the theoretical expected probability of such a k-mer arising (**Supplementary Table 1-2**). Due to the repetitive nature of the human genome, the corresponding actual probability of an avoidmer emerging in the human T2T genome is lower than the theoretical probability and this difference between expected and observed amplifies with increasing k-mer length and from $Z_2$ to $Z_3$ (**Figure 1b-c**). We conclude that there are fewer $Z_2$ and $Z_3$ avoidmers in the human genome than expected by a theoretical distribution and their number declines precipitously with increasing k-mer length.

**In the reference human genome every k-mer contains a Zimin word after 104 base-pairs**

As a next step, we wanted to identify the minimum length at which all k-mers cannot avoid $Z_3$ motifs. Specifically, we calculated the minimum length in the human T2T genome, after which all k-mers are not avoiding $Z_3$, $f\tilde{\,}(3)$ (Definition 1), which was calculated at 105 bp, with maximum Zimin avoidmer length of 104 bp emerging on chromosome 17 (**Figure 1d; Supplementary Table 3**). Theoretically, for an alphabet of size k the quantity $f(3,q)$ is bounded from above by $\sqrt{e}2^q(q + 1)! + 2q + 1$ (GöllerStefan 2019). Thus, $f(3,4)$ - and consecutively $f\tilde{\,}(3)$ - is *at most* 3,174 bp long (GöllerStefan 2019), which is significantly higher than the real $f\tilde{\,}(3)$ value of 105 bp. This discrepancy can be attributed to the fact that the human genome is a finite sequence which is highly repetitive and has a compositional bias with higher AT than GC content. Indeed, we find that Zimin avoidmers are significantly more GC rich than the genome average (**Figure 1e**). We conclude that every subsequence of the human genome, of at least 105 bps, contains a $Z_3$ Zimin pattern.

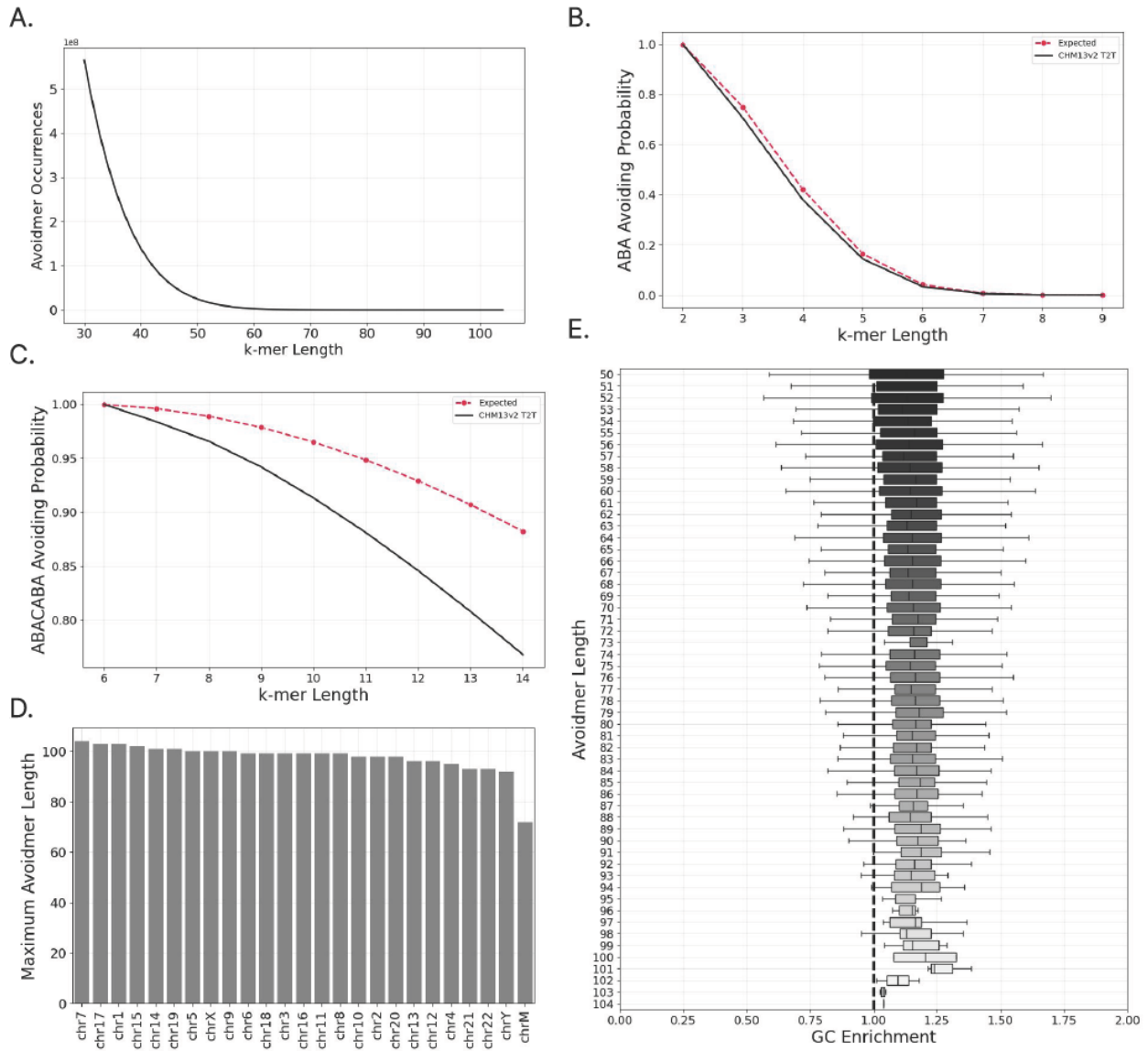

**Figure 1**: **Identification of Zimin words and Zimin avoidmers in the human genome. A)** Number of Zimin avoiding k-mers as a function of k-mer length, for k-mers ranging between 30 and 100bps. **B-C)** Expected and observed probability of a k-mer being Zimin avoiding for **B.** n=2 and **C.** n=3. **D).** Highest k-mer length at which avoidmers are observed in each chromosome in the human genome. **E)** Enrichment of GC content in avoidmers over the human genome average, for avoidmers of increasing length.

**Zimin avoidmer enrichment in specific genomic compartments**
Next, we examined if Zimin avoidmers are uniformly distributed in the human genome or if there are specific chromosomes and genomic subcompartments that display an excess. We first investigated if there exist significant differences in the distribution of avoidmers among the human chromosomes. We found that Zimin avoidmers of lengths of at least 60 bps have the highest genomic density in the Y chromosome, whereas shorter and longer Zimin avoidmers are more evenly spread between the remaining human chromosomes (**Figure 2a**).

We also investigated the density of avoidmers in various genomic subcompartments, such as genic, exonic and coding regions, as well as various pericentromeric and centromeric satellite regions, in order to identify potential hotspots of Zimin avoidmers. The vast majority of Zimin avoidmers of length equal or greater than 50 bps were located in coding sequence (CDS) regions (**Figure 2b**). When increasing the base pair threshold, Zimin avoidmers of at least 70bp long, were most enriched in classic human satellite hsat1B covering 2.6% of the satellite regions with an average density of 18,247.77 bp per Mb. The second most enriched genomic region were exonic regions displaying an averagedensity of 3,848.10 bp per Mb (**Figure 2b**). Despite the fact that satellite compartments are the main source of avoidmer elements, exonic and CDS loci are the third and second, respectively, most enriched genomic subcompartment in avoidmer population. Additionally, in the classical satellite hsat1B, the Zimin avoidmer density shows a sharp drop at 73bp, whereas the avoidmer distributions in exonic and coding regions exhibit a more gradual decline (**Figure 2b-c**). This could be attributed to the fact, that due to the repetitive nature of hsat1B the likelihood of $Z_3$ motif eventually emerging increases disproportionally at higher k-mer lengths. The Y chromosome is the most rich chromosome in hsat1B regions and we find that a small subset of Zimin avoidmers originate in these highly repetitive regions (**Supplementary Figure 1**). Most of the Zimin avoidmer satellite sequences detected in hsat1B are exactly the same sequence of 73bp long, appearing 4,212 times (**Supplementary Table 4**). Interestingly, the second most frequent Zimin avoidmer in hsat1B is also 73bp long and has 277 occurrences, and differs from the first most frequent sequence by a single bp substitution (**Supplementary Table 4**). These two sequences explain the sharp drop in Zimin avoidmer density in the classical satellite hsat1B that we observed previously. We conclude that Zimin avoidmers have an inhomogeneous distribution in the human genome, and are most enriched in CDS and hsat1B regions.

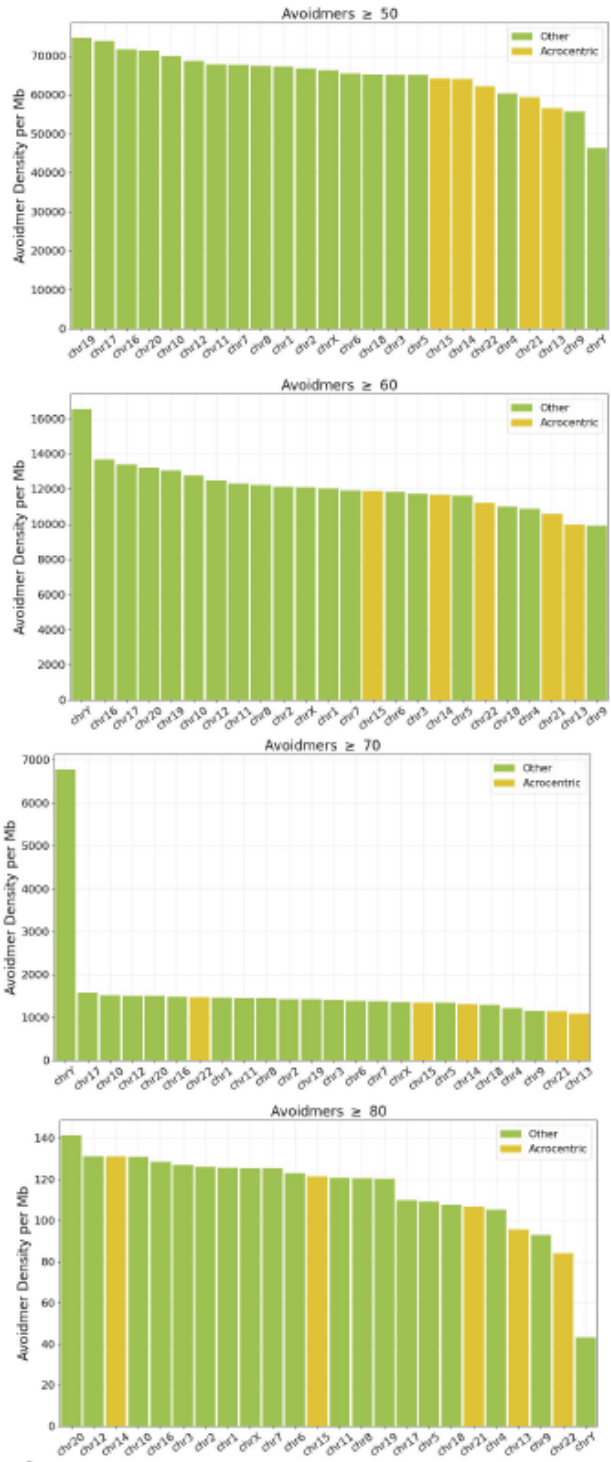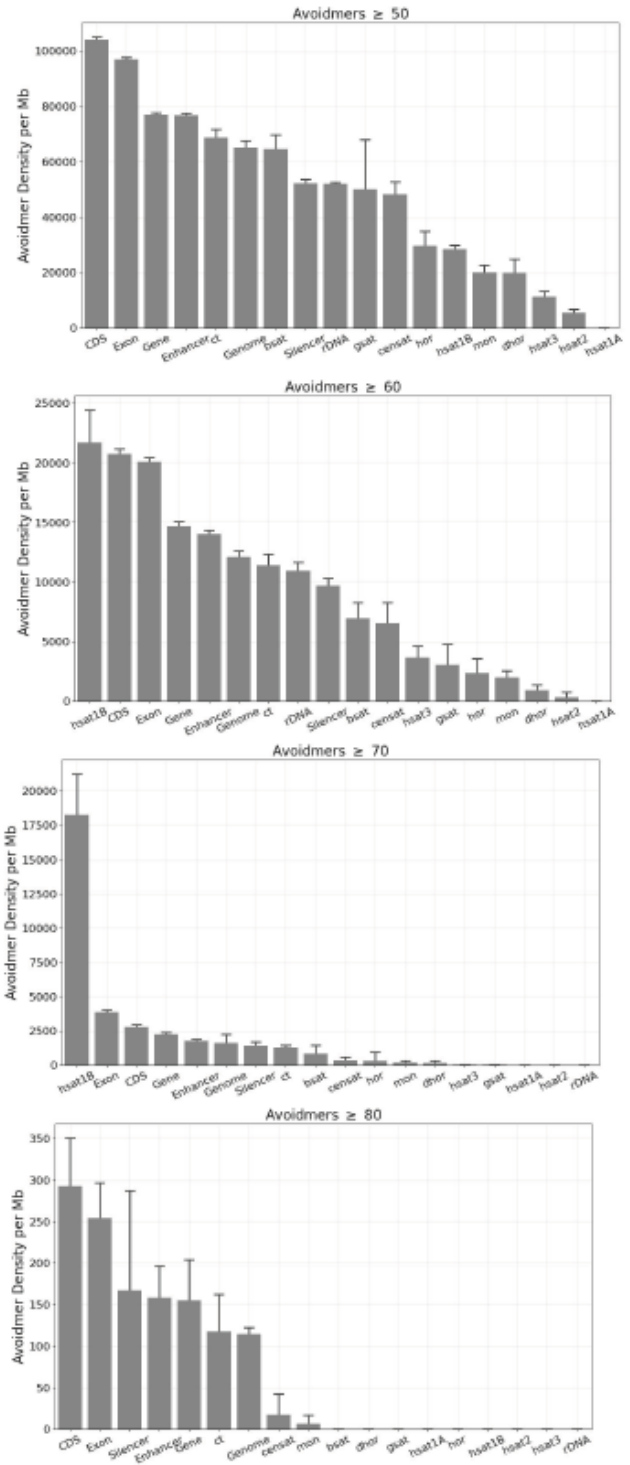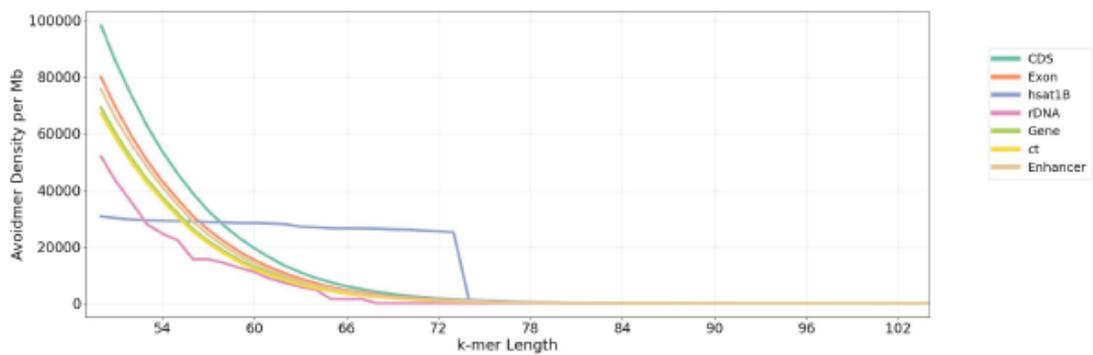

**Figure 2**: **Zimin avoidmers are inhomogeneously distributed in the human genome. A)** Avoidmer density per Mb across various human T2T chromosomes for k-mer lengths of at least 50, 60, 70 and 80 bp. **B)** Avoidmer density per Mb in human pericentromeric and centromeric satellite and genic subcompartments for k-mer lengths of at least 50, 60, 70 and 80 bp, respectively. **C)** Zimin avoidmer k-mer density as a function of k-mer length across the genomic compartments.

**Zimin avoidmers are associated with increased sequence diversity**
Next, we binned the genome in non-overlapping 50 kilobase (kB) windows to examine how the k-mer diversity influences the number of avoidmers detected. In each bin we counted the number of unique k-mers detected for eight bps k-mer length. Then, we quantified the correlation between the avoidmer occurrences and distinct k-mer sequences detected in each bin. We observe that the Zimin avoidmers with the largest number of occurrences are primarily observed in bins in which there is large k-mer diversity (**Figure 3a**). These results are also consistent in genic and CDS compartments (**Figure 3b**). However, we find that a small subset of Zimin avoidmers are found in bins that have low k-mer diversity and these originate from satellite repeat regions, such as hsat3 (**Figure 3c**). Next, to examine potential repetitive patterns in Zimin avoidmers, we investigated the density of subsequences that are an exact instance of $Z_2 = aba$ in Zimin avoidmers (Rorabaugh 2015). We observe that the density of ABA sequences in Zimin avoidmers follows a normal distribution; however with an elongated tail representing a small subset of Zimin avoidmers that have high density of ABA sequences (**Figure 3d**). Interestingly, the average ABA density was anti-correlated with the number of distinct k-mers present in each Zimin avoidmer (**Figure 3e**), indicating that, on average, Zimin avoidmers without ABA words have the highest k-mer diversity. In fact, on average, we anticipate the same effect in any k-mer. We conclude that Zimin avoidmers are associated with loci that have increased sequence diversity in the human genome.

We were interested to determine how long, on average, a uniformly randomly generated k-mer would be before it encounters an instance of $Z_3$. We simulated a simple random process which constructs a k-mer by choosing at random - with probability 0.25 - a nucleotide from the alphabet {a, g, c, t}. This random process continues until the resulting sequence encounters an instance of $Z_3$. We repeated this process 50,000 times and documented the resulting Zimin avoidmers for each of the random experiments. We estimated the average value of maximum length of randomly generated Zimin avoidmers being equal to 26.6bp (**Supplementary Figure 2a**). However, as quite a few outliers reach higher lengths (**Supplementary Figure 2b**). Finally, we examined the relationship of the lengths of the resulting sequences with the ABA density. The latter indicated a logarithmic relationship with the maximum lengths, suggesting that a small increase in the total length induces an exponential increase in the subsequences that are instances of $Z_2$ (**Supplementary Figure 2c**). Although this random process relies on basic assumptions that do not apply to regular genomes, such as nucleotide positional and local independence, it enables us to quantify the inherent difficulty of constructing arbitrarily long Zimin avoiding k-mers by relying purely on chance.

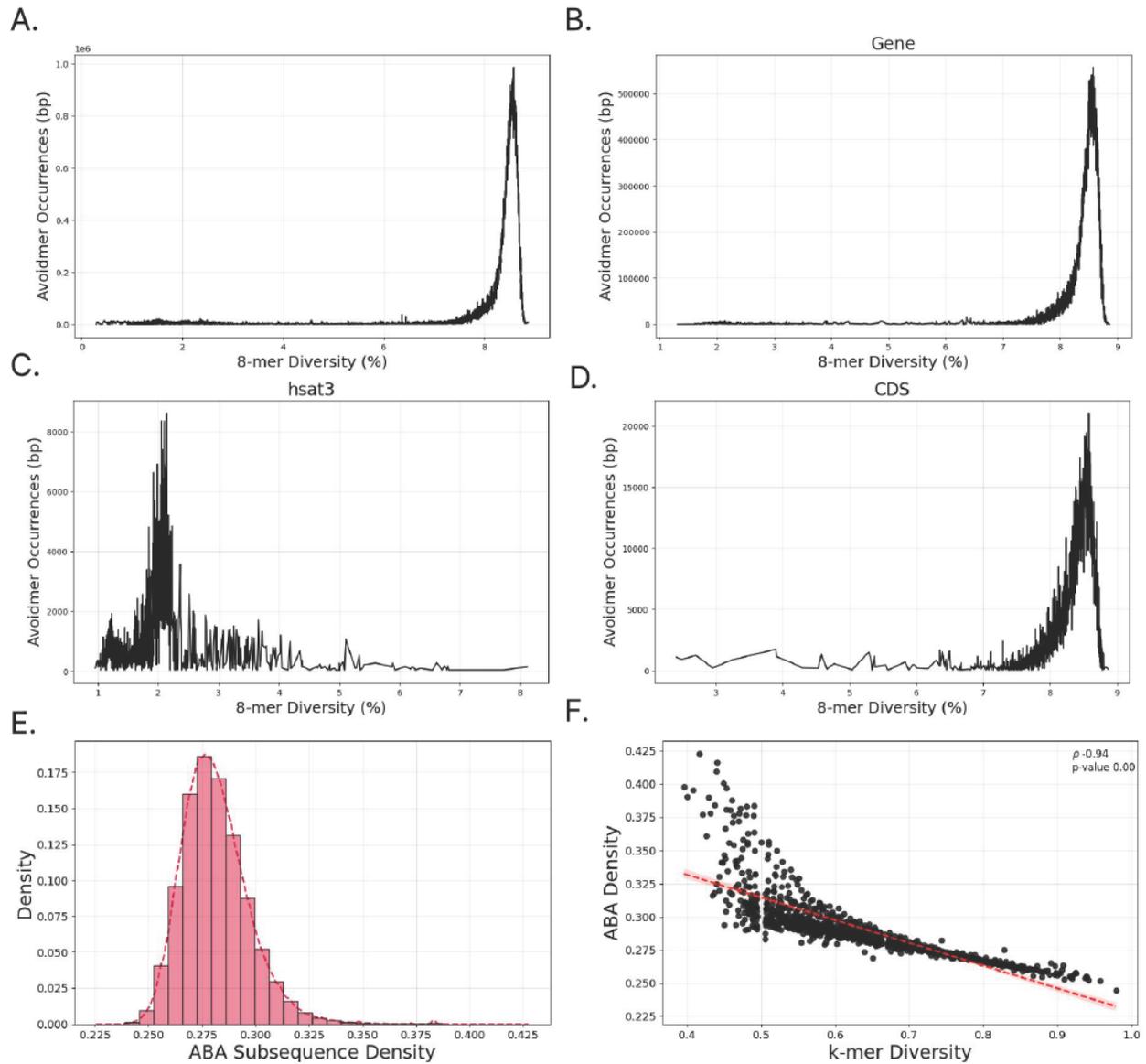

**Figure 3: Characterization of the sequence diversity in avoidmers. A.** Percentage of possible 8-mers and avoidmer occurrences present in 50kB genomic bins. **B-D.** Percentage of possible 8-mers and avoidmer occurrences present in 50kB genomic bins in: **B.** Genes, **C.** Hsat3 compartments**, D.** CDS regions. **E.** ABA sub-sequence density in avoidmer sequences. **F.** ABA density in avoidmers as a function of average k-mer diversity in genomic bins.

**In silico saturation and germline mutagenesis of Zimin avoidmers in the human genome**

We investigated if the pattern of being a Zimin avoidmer is resistant to the introduction of single bp indels and substitutions, a property that we hereafter term invariance. We simulated all possible one bp indels and substitutions at each Zimin avoidmer. Zimin avoidmers exhibited a higher rate of invariance for deletions, followed by insertions, and lastly, substitutions (**Figure 4a**). On average, when increasing the sequence length by randomly inserting a random letter from the nucleotide alphabet the probability of the Zimin property diminishes, due to the fact that it is less likely for a longer sequence to be a Zimin avoidmer (**Figure 1a**). Thus, it is more likely for deletions to maintain that initial invariance, in contrast to random insertions or substitutions (**Figure 4a**). Furthermore, due to the fact that for every Zimin avoidmer by definition, each subsequence must also be a Zimin avoidmer, deletions at the end of the k-mer sequence, must inevitably maintain the Zimin avoidmer property (**Supplementary Figure 3**).

Subsequently, we used germline variants from The Genome Aggregation Database (gnomAD) (Karczewski et al. 2020) and examined the positioning of Zimin avoidmers relative to substitutions and indels. We constructed a 1kB window centered at the germline mutations and counted the number of Zimin avoidmer occurrences in each position. We observe a strong depletion of Zimin avoidmers at the proximal region surrounding insertions and deletions (**Figure 4b**). Interestingly, we do not find strong enrichment or depletion patterns for substitutions (**Figure 4b**). Due the complex nature of sequences that avoid $Z_3$ patterns, we postulated that the strong depletion signal of indels originates from absence of Short Tandem Repeats (STRs) at Zimin avoidmers, which are highly enriched for indels (Georgakopoulos-Soares et al. 2022; Ananda et al. 2013) and depleted in $Z_3$ avoidmers. We examined the relative positioning of STRs in respect to Zimin avoidmer sequences. We found that, as expected, STRs are depleted in the proximity of Zimin avoidmer sequences, particularly in CDS and genic compartments (**Figure 4c**). This result coincides with our intuition suggesting that due to the inherent complexity of CDS compartments, Zimin avoidmers are most enriched in those regions and are depleted from the repetitive STR loci, impacting their mutation rates.

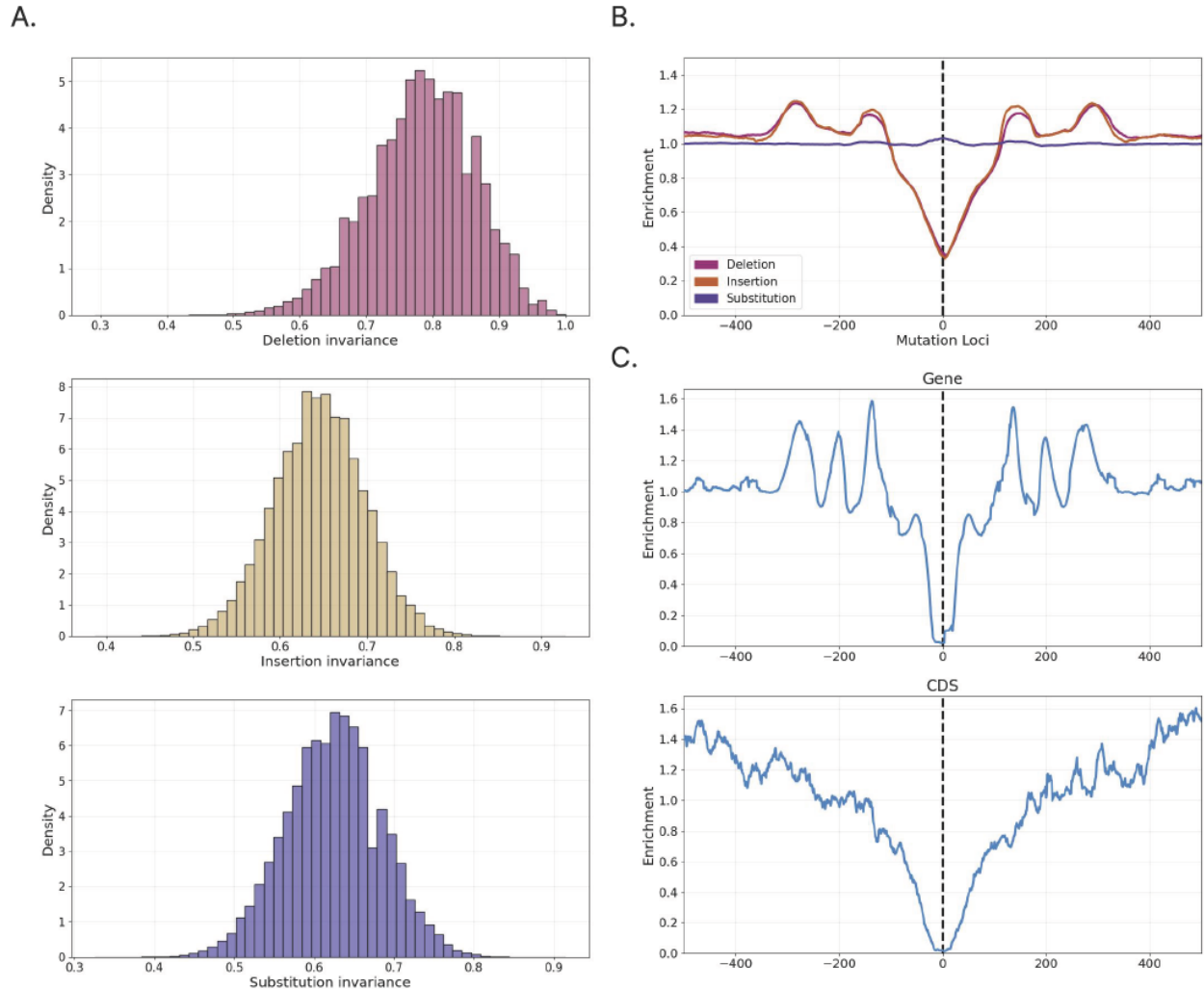

**Figure 4**: **A)** Histogram of invariance of avoidmers under mutational transformations. **B)** Relative positioning of avoidmers in relationship to substitutions, deletions and insertions across 1kB window. **C)** Relative positioning of STRs in relationship to genic and CDS avoidmers, respectively.

**Identification of Zimin words and Zimin avoidmers across model organism genomes**
Next, we examined how our findings translate in other organismal genomes. We selected a set of reference genomes from model organisms, spanning all three domains of life (**Table 1**), and estimated the k-mer length after which every k-mer contains a Zimin word. We observe that the highest Zimin avoidmer length is observed for *S. cerevisiae* for k-mer length of 115 bps, whereas the lowest is for *S. aureus* at 86 bps (**Figure 5a, Table 1**). We also find that the $Z_3$ avoidmer density of Zimin avoidmers of at least 50bp length, varies substantially between the examined organisms, ranging between 180,905.07 and 58,493.34 $Z_3$ avoidmers per mB in *E. coli* and *D. rerio* respectively. We investigated the expected theoretical probability and observed probability of observing Zimin avoidmers in each of the model organisms. We find that the largest discrepancies between expected and observed probabilities are in eukaryotic organisms including *G. galus*, *D. rerio*, *D. melanogaster*, *S. cerevisiae* and *C. elegans*, whereas the smallest differences are observed in prokaryotes, namely *S. aureus*, *K. pneumoniae* and *E.coli* (**Figure**

**5c-d**). This can be likely explained by the more repetitive and nucleotide imbalanced genomes of eukaryotes compared to prokaryotes.

**Zimin avoidmers are inhomogeneously distributed across model organismal genomes**
For each examined organism we separated their genome in genic and intergenic regions and examined the percentage of Zimin avoidmers in each. We find that across organisms, Zimin avoidmers have more occurrences in genic than intergenic regions (**Figure 5e**), results that are consistent with our observations for the human genome. When further separating the genomic compartments in genic, exonic and cds regions we find that the Zimin avoidmer density varies substantially both between sub-compartments and between the species (**Figure 5f**). We conclude that across most of the studied organisms, Zimin avoidmers are most enriched in genic and CDS regions of the genome, particularly in the case of organisms of eukaryotic origin.

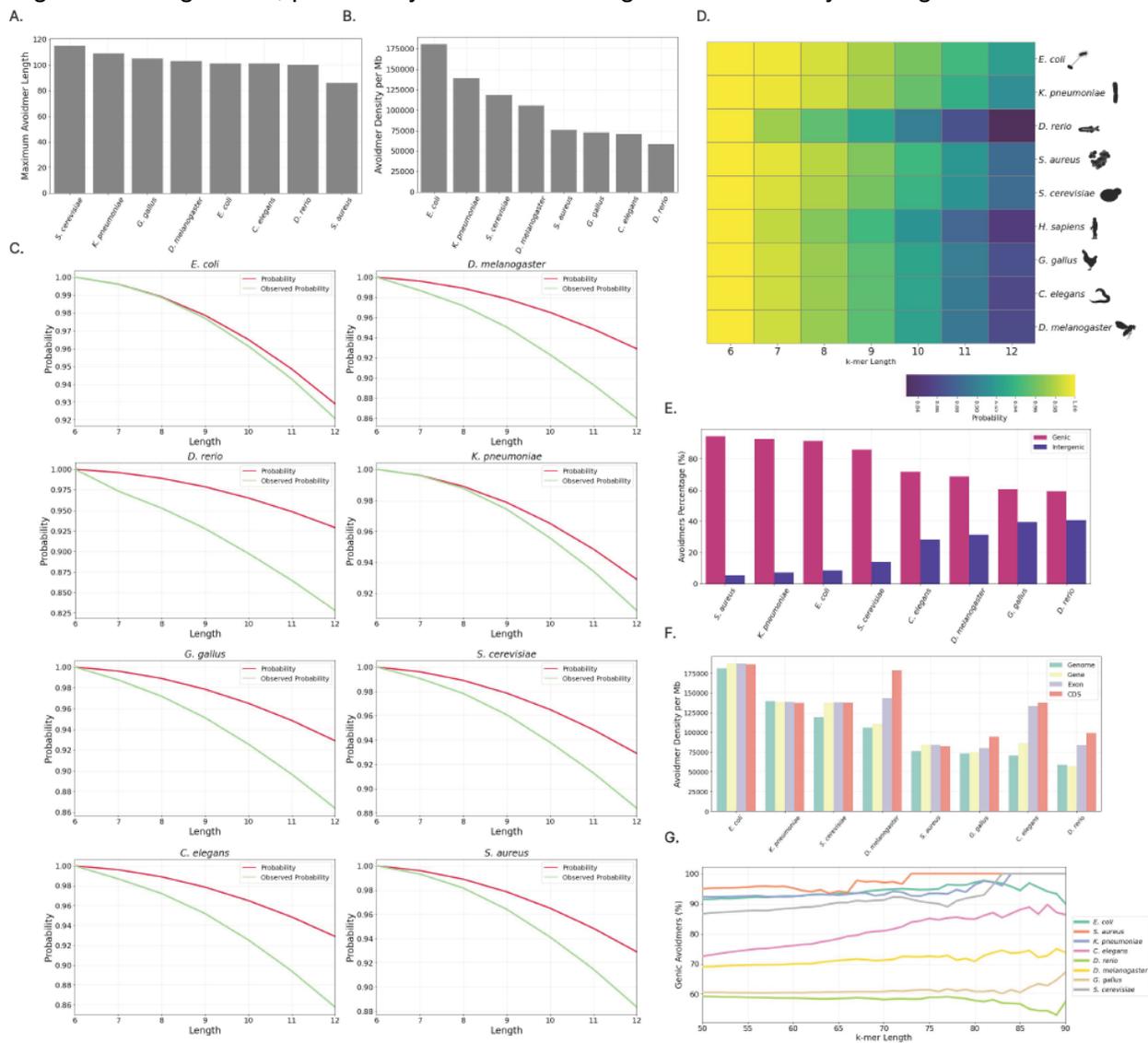

**Figure 5**: **Identification of avoidmers across model organisms. A)** Density of Zimin avoiding k-mers across model organisms. **B)** Highest k-mer length at which avoidmers are observed in each chromosome in the human genome. **B)** Zimin avoiding k-mer density across human

chromosomes. **C)** Expected and observed probability of k-mer being Zimin avoiding across model organisms. **D)** Observed probability of a k-mer being avoidmer across model organisms. **E)** Avoidmer percentage across genic and intergenic regions for each model organism. F) Zimin avoidmer density across genomic subcompartments for each model organism. **F)** Avoidmer percentage across compartments for each model organism. **G)** Zimin avoiding k-mer percentage for increasing k-mer length for each of the under study model organisms.

**Table 1:** Highest k-mer length at which avoidmers are observed in each model organism's genome.

| Species | Avoidmer threshold (bp) |
|---|---|
| *Saccharomyces cerevisiae* | 115 |
| *Klebsiella pneumoniae* | 109 |
| *Gallus gallus* | 105 |
| Homo sapiens | 104 |
| *Drosophila melanogaster* | 103 |
| *Caenorhabditis elegans* | 101 |
| *Escherichia coli O157:H7* | 101 |
| *Danio rerio* | 100 |
| *Staphylococcus aureus* | 86 |

**Discussion**

We have characterized for the first time Zimin words and Zimin avoidmers in the genomes of nine organisms, including the human genome and the genomes of multiple model organisms. This is a mathematical concept, with significant previous theoretical research (Choffrut and Karhumäki 1997; Lothaire 2014; GöllerStefan 2019), that is applied for the first time in a genomics setting. By construction, genomic sequences which avoid Zimin patterns encapture a higher level of grammar complexity, since they are bound to avoid certain repetitive patterns in order to maintain their Zimin avoidability property. Furthermore, on sufficiently large genomic sequences, the unavoidability theorem suggests that the emergence of such patterns is unavoidable (Zimin 1984). We observe that Zimin avoidmers disappear from organismal genomes at lower k-mer lengths than expected from the theoretical upper limit (**Table 1**). This is likely driven by the repetitive nature of organismal genomes and the uneven frequency of nucleotides. We also observe that Zimin avoidmers are inhomogeneously distributed in organismal genomes. In the human genome Zimin avoidmers are over-represented in the Human Satellite 1B compartments and in coding sequences.

Zimin avoidmer sequences, by construction, do not contain any STR as a proper subsequence. Thus, these artificial nucleotide regions encapture a higher level of irregularity which is interrupted by canonical patterns of DNA, such as tandem repeats. The total density of Zimin avoidmers is more pronounced in prokaryotic genomes rather than eukaryotic genomes which could potentially be related to the depletion of microsatellite sequences in certain bacteria phylums (Chantzi and Georgakopoulos-Soares 2024).

Zimin sequences and their counterparts, the Zimin avoidmers, are of particular interest and hold potential utility in bioinformatics. Their inherent symmetry and self-similarity could link them with non-canonical DNA conformations also known as non-B DNA motifs (Wang and Vasquez 2023), as these patterns exhibit both direct and mirror-repeat symmetries. This raises the question of whether the presence or absence of such properties in k-mers plays a crucial role in the formation of various DNA secondary structures across different genomes. We believe that the characterization of Zimin words provides a new lens through which one can view genomes using homomorphism embeddings of various patterns. The constraints associated with their repetitive nature and unbalanced nucleotide frequencies. Future studies could implement additional concepts from algebraic combinatorics on words in genomics to acquire a deeper understanding of rules of genomic grammar. Finally, future work is required to further examine Zimin words in organismal genomes and their emergence or loss through population variants as well as potential applications and development of tools based on them.

**Methods**

**Data Retrieval**

We downloaded the reference human genome assembly T2T-CHM13v2.0. Associated files including gene annotation and comprehensive centromere/satellite repeat annotation files were downloaded from https://github.com/marbl/CHM13. The gene annotation GFF file was downloaded from https://ftp.ncbi.nlm.nih.gov/genomes/all/GCF/009/914/755/GCF_009914755.1_T2T-CHM13v2.0/GCF_009914755.1_T2T-CHM13v2.0_genomic.gff.gz and a more comprehensive centromere/satellite repeat annotation file was derived from https://s3-us-west-2.amazonaws.com/human-pangenomics/T2T/CHM13/assemblies/annotation/chm13v2.0_censat_v2.0.bed. We also downloaded the following complete genomes: *Haloferax volcanii DS2* from https://ftp.ncbi.nlm.nih.gov/genomes/all/GCF/000/025/685/GCF_000025685.1_ASM2568v1/GCF_000025685.1_ASM2568v1_genomic.fna.gz, *Halobacterium salinarum* from https://ftp.ncbi.nlm.nih.gov/genomes/all/GCF/004/799/605/GCF_004799605.1_ASM479960v1/GCF_004799605.1_ASM479960v1_genomic.fna.gz, *C. elegans* from https://hgdownload.soe.ucsc.edu/goldenPath/ce11/bigZips/ce11.fa.gz, *D. melanogaster* from https://hgdownload.soe.ucsc.edu/goldenPath/dm6/bigZips/dm6.fa.gz, *G. gallus* from https://hgdownload.soe.ucsc.edu/goldenPath/galGal6/bigZips/galGal6.fa.gz, *S. cerevisiae* from https://hgdownload.soe.ucsc.edu/goldenPath/sacCer3/bigZips/sacCer3.fa.gz, Zebrafish from https://hgdownload.soe.ucsc.edu/goldenPath/danRer11/bigZips/danRer11.fa.gz, *Klebsiella*

*pneumoniae* subsp. pneumoniae HS11286 from https://ftp.ncbi.nlm.nih.gov/genomes/all/GCF/000/240/185/GCF_000240185.1_ASM24018v2/GCF_000240185.1_ASM24018v2_genomic.fna.gz, *Escherichia coli* O157:H7 from https://ftp.ncbi.nlm.nih.gov/genomes/all/GCF/000/008/865/GCF_000008865.2_ASM886v2/GCF_000008865.2_ASM886v2_genomic.fna.gz, *Staphylococcus aureus* subsp. aureus NCTC 8325 from https://ftp.ncbi.nlm.nih.gov/genomes/all/GCF/000/013/425/GCF_000013425.1_ASM1342v1/GCF_000013425.1_ASM1342v1_genomic.fna.gz. We also downloaded GTF files for gene annotation for each of these organismal genomes. Single nucleotide substitutions, insertions and deletions were downloaded for the Human Pangenome Reference Consortium from year one data from https://s3-us-west-2.amazonaws.com/human-pangenomics/pangenomes/freeze/freeze1/minigraph-cactus/hprc-v1.1-mc-chm13/hprc-v1.1-mc-chm13.vcfbub.a100k.wave.vcf.gz.

**Identification of Zimin avoiding sequences**

To extract the *Zimin avoiding sequences* across the different kmer lengths, we recursively constructed a backreferencing regular expression pattern matching using Python, as shown in Supplementary Table 2. For each chromosome, we scan the sequence from left to right, reading chunks of k-mers of at least *L* bp long. For each of these chunks, the Zimin avoidance property is examined. We are attempting an extension by reading one more base pair at a time, and testing Zimin avoidance. We continue this extension process, until one additional bp breaks Zimin avoidance. Before we exit the loop, we save the k-mer without the final base pair, as Zimin avoidmer. The process continues until we have exhausted the whole genomic sequence. Overlapping sequences are merged using pybedtools (Dale, Pedersen, and Quinlan 2011) before analyzing and reporting the genomic sub-compartment densities for the studied organisms. The GitHub link to the code that performs the detection of Zimin avoidmers is provided in the relevant section.

| Length | Regular expression pattern | Zimin avoidmer |
|---|---|---|
| n=1 | ([agct]+) | $Z_1 = a$ |
| n=2 | ([agct]+)([agct]+)\1 | $Z_2 = aba$ |
| n=3 | ([agct]+)([agct]+)\1([agct]+)\1\2\1 | $Z_3 = abacaba$ |
| n=4 | ([agct]+)([agct]+)\1([agct]+)\1\2\1([agct]+)\1\2\1\3\1\2\1 | $Z_4 = abacabadabacaba$ |

**Table 2: Regular expression pattern used to extract Zimin avoiding sequences**

**Estimation of expected and observed Zimin word frequencies**

The probability of observing a Zimin word within a particular organismal genome G was estimated using the total number of avoidmers for a particular k-mer length divided by the total number of k-mer occurrences for that particular genome, i.e.

$$P_G(Z_3) = \frac{total\ avoidmers}{total\ number\ of\ k-mers}.$$

The k-mer occurrences were calculated using Jellyfish (Marçais and Kingsford 2011) for k-mer length between 1 and 13bp for each organismal genome in this study.

The expected probability was calculated by recursively generating the nucleotide k-mer tree while pruning sequences that are not Zimin avoidmers. At each recursive step, the total number of k-mers that belong to the avoidmer class were calculated and subsequently divided by the total number of k-mers for that particular genome to derive the expected probability.

**Examination of avoidmers in genomic compartments**
The density of Zimin avoidmers in genic, exonic, 5' UTR, 3' UTR, and CDS regions as well as in centromeric and pericentromeric repeat regions was calculated using publicly available GTF and GFF coordinate files. We used compartments annotated as "transcript" from GTF files and "gene" from GFF files to find genic and intergenic areas. A Zimin avoidmer was considered "genic", if it had at least 1bp overlap with a genic region from the provided GFF/GTF file. Additionally, for the prokaryotic genomes we used agat ("NBISweden/AGAT: AGAT-v1.4.1," n.d.) to fill up the missing exon annotations. For each annotated subcompartment, overlapping compartments of the same type were merged and expanded into non-overlapping, mutually disjoint sets of coordinates. These coordinates were then used to estimate the total coverage per Mb by calculating the total number of bps of avoidmers divided by the total compartment length of the previously merged genomic regions.

**Quantification of mutation frequency in the vicinity of Zimin words**
We generated a 500 bp window upstream and downstream from each Zimin word's genomic coordinates center and estimated the mutation rate at each bp at each bp. The enrichment was calculated as the number of mutation occurrences at a position over the mean number of occurrences across the window.

**Quantification of k-mer diversity across the human genome**
Given a k-mer w, we define the subsequence $w[i:i+l]$ of w, as the $l$-mer that starts at position $0 \leq i \leq |w| - l$ and ends at position $i + l - 1$ (inclusive) with $1 \leq l \leq |w|$. The diversity of k-mer w and subsequently avoidmer sequences for a given length $L$ was determined as the ratio of the total number of unique subsequences to the total number of subsequences of the given length $L$, i.e.

$$\text{k-mer diversity} = \frac{|\{w[i:i+l]:\ 0 \leq i \leq |w|-l+1\}|}{|w|-l+1},$$

where by |w| denote the length of the k-mer.

To investigate the relationship of avoidmers to high complexity genomic areas, we partitioned each chromosome of the *Homo sapiens* T2T assembly into non-overlapping, consecutive 50kB window bins. For each of the bins, we evaluated the associated k-mer diversity as defined above. Subsequently, we calculated the number of Zimin avoidmers in each genomic bin. Then we compared the association between the k-mer diversity and the total number of occurrences of Zimin avoidmers.

**Monte-Carlo simulations to characterize Zimin avoidmer properties**

We simulated a random process by which we generated random Zimin avoidmer sequences. We constructed a Zimin avoidmer by repeatedly sampling one nucleotide from the nucleotide alphabet {'a', 'g', 'c', 't'} at random. At each step, the resulting sequence was evaluated to satisfy the Zimin avoiding criteria. If at any step, the new nucleotide was violating the Zimin avoiding property, we saved the length of the previous Zimin avoiding sequence. The process was repeated 50,000 times. The resulting sample data were stored in a pandas dataframe and were processed to generate the simulation figures. Additionally, by repeatedly resampling from the simulated dataset with increasing sample size, initiating at 50 and reaching 5,000, with 10 sample increments between iterations, by the Law of Large numbers, we were able to asymptotically evaluate the average length of avoidmer sequences generated by this simple stochastic process. However, note that the generated average length, is of course, bounded to this simple stochastic experiment, and cannot be generalized to sequences not generated by this exact random process. Finally, the ABA subsequence density was evaluated for each of the generated Zimin avoidmer sequences, revealing a logarithmic relationship between the length of the avoidmer sequence and the subsequence ABA density.

***In silico* saturation mutagenesis in Zimin avoidmers**

To determine the mutational invariance of each Zimin avoidmer, we used a custom Python script which exhaustively generates all possible substitutions, deletions, and insertions and examined if the avoidmer property of the resulting k-mer remained invariant. For each Zimin avoidmer we saved the proportion of possible variants that preserved the Zimin avoiding property to the total number of possible variants, separately for insertions, substitutions and deletions. Additionally, for each avoidmer we examined the proportion of mutations for which the resulting mutated sequence remained an avoidmer, which we termed invariance. Because each Zimin avoidmer has varying length, for each sequence, we partitioned the k-mer length into 22 mutually exclusive bins, and for each bin, the average invariance was evaluated as the average number of mutations that kept the Zimin property.

**Examination of germline variants at Zimin avoidmers**

We used the germline mutations from The Genome Aggregation Database (gnomAD) (Karczewski et al. 2020). Each mutation was classified into one of the following categories: Substitution, Insertion, Deletion and MNP. The allele frequency was not taken into account. Subsequently, a 1kB window was constructed, centered at each of the classified mutational loci. Utilizing the pybedtools package and the intersect function, for each of the previously constructed intervals, we used the BED coordinates of Zimin avoidmers, and mapped each of them at each window, whenever an intersection occurred. Finally, we calculated for each position across the

1kB window the number of Zimin avoidmer bps relative to the mutational loci. Finally we plotted the enrichment of Zimin avoidmers across the 1kB window using the matplotlib and Seaborn Python libraries (Waskom 2021).

**Zimin avoidmer distribution relative to Short Tandem Repeats**
To determine the frequency of Zimin avoidmers relative to STRs, we extracted STRs for human T2T genome chm13v2 as described in (Chantzi and Georgakopoulos-Soares 2024). Using the aforementioned procedure, we constructed 1kB intervals centered around the Zimin avoidmers, and examined the number of occurrences of STR bps across the 1kB window, across all Zimin avoidmer loci. We calculated the total occurrences at each position resulting and divided by the mean occurrences across the window to evaluate the enrichment of STR at each position relative to Zimin avoidmers.

**ABA Density**
The ABA density of a k-mer w, is defined as the ratio of the total number of subsequences of w that are an instance of ABA $Z_2$ motif to the total number of subsequences of w (Rorabaugh 2015), i.e.
$$\lambda(w) = 2\frac{\#\{subsequences\ that\ are\ instance\ of\ Z_2\}}{N(N+1)},$$

where $N = |w|$. This formula was implemented in a custom C++ script to extract the ABA density for each of the extracted Zimin avoidmer sequences. Note that the function $\lambda(w)$ is bounded above by 1. In fact, the upper bound can be improved. By using the observation that the *N* mononucleotides and *N-1* dinucleotide subsequences cannot possible form an instance $Z_2$. Thus, by subtracting these $2N - 1$ mononucleotide and dinucleotide subsequences, we conclude that:
$$\lambda(w) \leq 1 - 2\frac{2N-1}{N(N+1)}.$$

**Zimin avoiding patterns do not encounter Tandem Repeats**
A tandem repeat is any k-mer which can be written in the form $x^m = x\ldots x$ *m* times. The *x* is often referred to as the *consensus sequence* of tandem repeat and number $m$ as the consensus repeats. A microsatellite is defined as any tandem repeat with $|x| \leq 9$. We claim that any tandem repeat of at least 4 consensus repeats encounters a $Z_3$ pattern when the consensus sequence is at least 2bp long.
For mononucleotides with consensus motif $x$ with $|x| = 1$ and $m \geq 7$, $x^m$ encounters the pattern $Z_3$ for $f(a) = f(b) = f(c) = x$, as $f(Z_3) = \boldsymbol{f(abacaba)} = \boldsymbol{x^7}$ is a subsequence of $x^m$.
For $|x| \geq 2$ and $m \geq 4$, we can expand $x$ in the form $x = yz$, and, consequently $x^m = (yz)^m$. By definining the homomorphism $f$ with $f(a) = y$, $f(b) = f(c) = z$, we note that:
$$\begin{aligned}f(Z_3) &= \boldsymbol{f(abacaba)}\\ &= \boldsymbol{f(a)f(b)f(a)f(c)f(a)f(b)f(a)}\\ &= yzyzyzy = (yz)^3y.\end{aligned}$$
But $f(Z_3) = (yz)^3y$ is a subsequence of $x^m$, as

$$x^m = (yz)^m = (yz)^3 yz(yz)^{m-4} = f(Z_3)z(yz)^{m-4}.$$

Thus $x^m$ encounters $Z_3$ for all cases where $|x| = 1 \ \& \ m \geq 7$ and $|x| \geq 2 \ \& \ m \geq 4$. We conclude that avoidmers cannot coincide with the majority of tandem repeats loci.

**Supplementary Material**

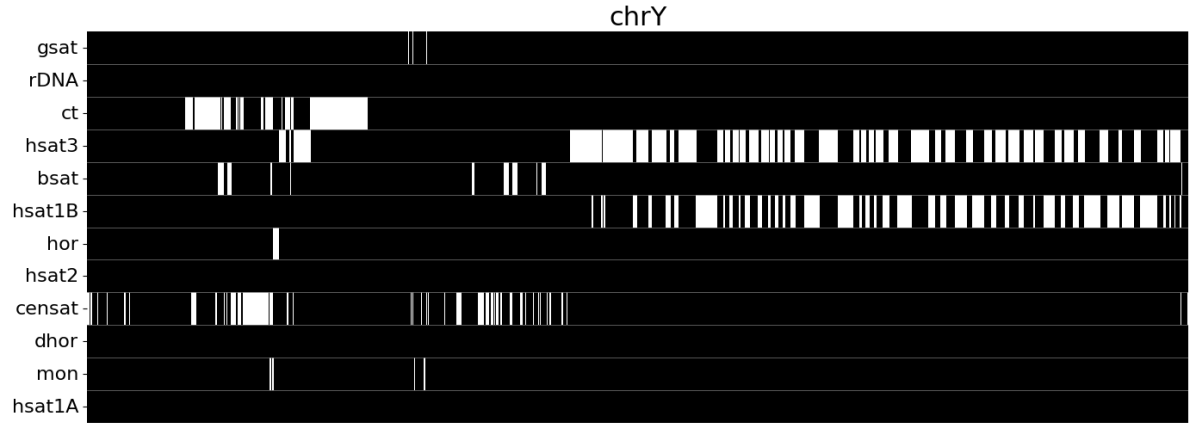

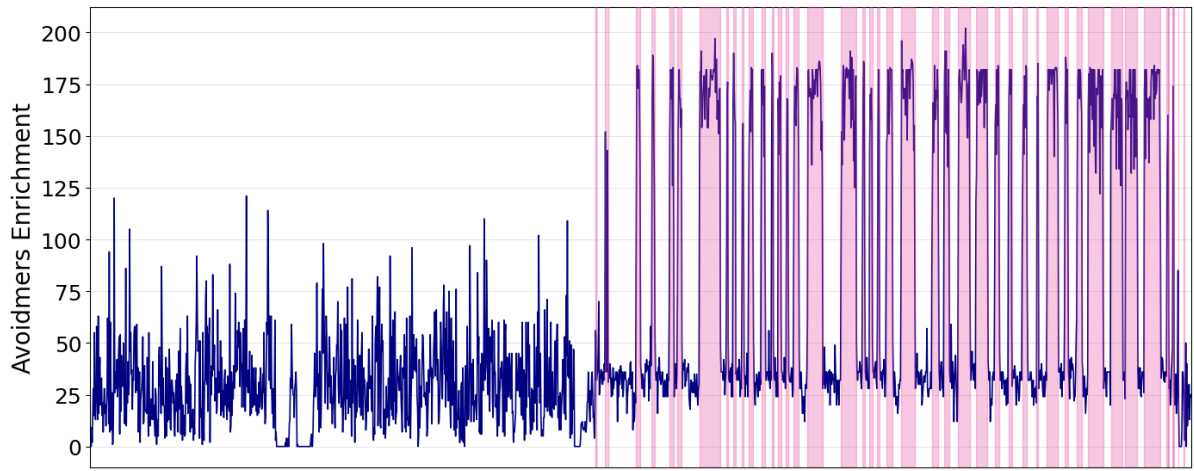

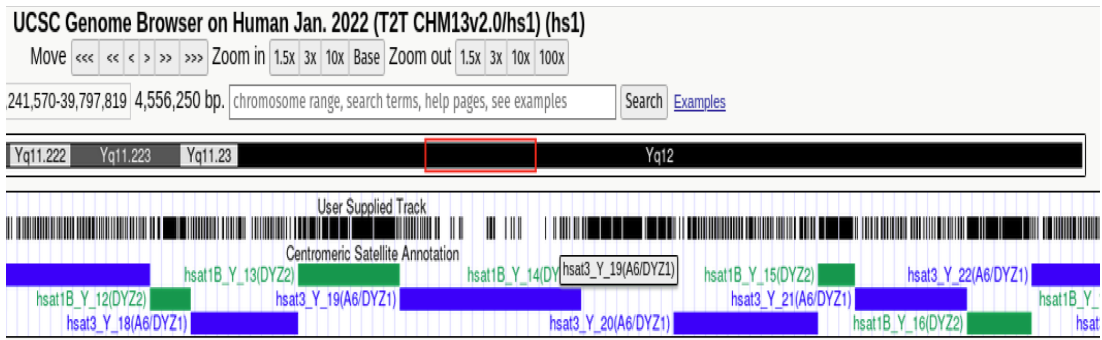

**Supplementary Figure 1: Avoidmer density across the Y chromosome.** The highlighted regions correspond to the classical human satellite region hsat1B where the vast majority of avoidmers are located. Furthermore, using the UCSC Genome-Browser, we visualized the distribution of Zimin avoiding sequences of at least 70bp in a section of the satellite-heterochromatic area of Y chromosome. The densities of Zimin avoidmer sequences of at least 70 bp long, appear denser in the hsat1B compartments rather than the hsat3 satellite regions.

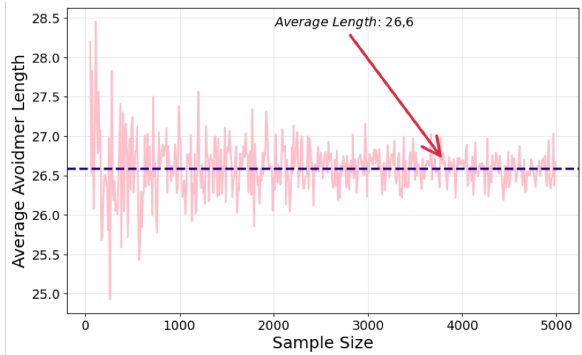
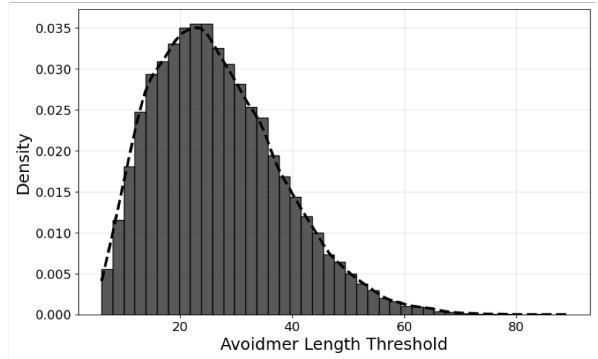
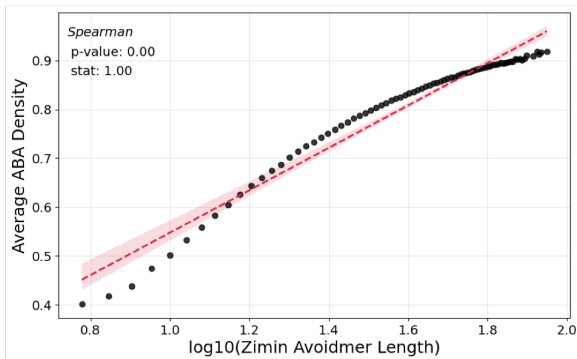

**Supplementary Figure 2: Examination of Zimin avoidmer length and composition distribution. A)** Average avoidmer length threshold approximation by increasing sample size, following the law of large numbers. Random sampling with replacement was used, with sample sizes ranging between 50 and 5,000, with 10 sample increments between iterations. **B)** Avoidmer length threshold distribution from random equidistributed simulations using the nucleotide alphabet. **C)** Average ABA Density is exponentially increased with Zimin avoidmer length.

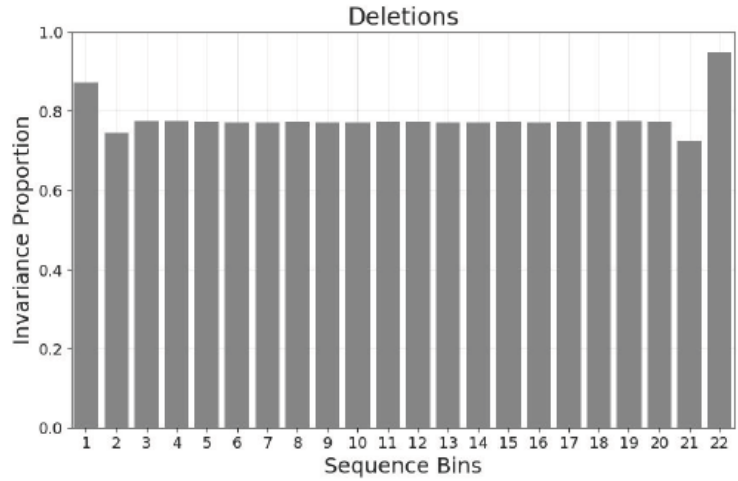
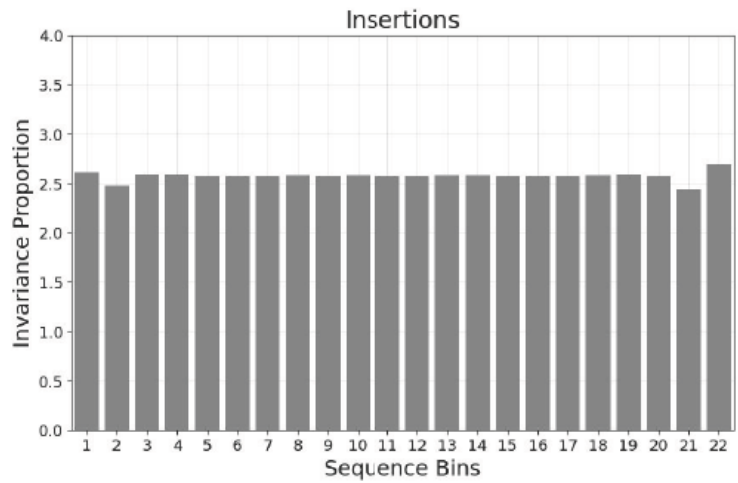
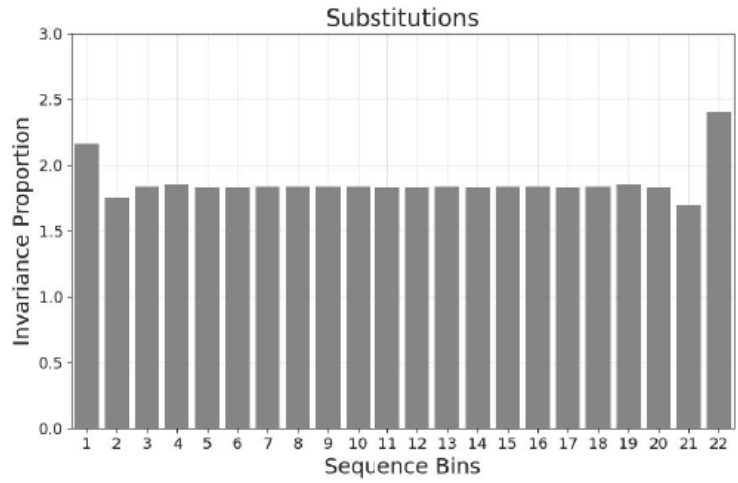

**Supplementary Figure 3: Mutational Invariance for Substitutions, 1bp Insertions, 1bp Deletions.** Each Zimin Avoidmer was partitioned into twenty mutually exclusive bins and for each bin, the invariance potential for each of the mutations was estimated.

**Supplementary Table 1: ABA Avoiding Expected Probabilities**

| K-mer Length | Total K-mers | ABA Avoidmers | Expected Probability (%) |
|---|---|---|---|
| 1 | 4 | 4 | 100.00 |
| 2 | 16 | 16 | 100.00 |
| 3 | 64 | 48 | 75.00 |
| 4 | 256 | 108 | 42.1875 |
| 5 | 1,024 | 168 | 16.40625 |
| 6 | 4,096 | 168 | 4.1015625 |
| 7 | 16,384 | 96 | 0.5859375 |
| 8 | 65,536 | 24 | 0.03662109375 |
| 9 | 262,144 | 0 | 0,00 |

**Supplementary Table 2: ABACABA Avoiding Expected Probabilities**

| K-mer Length | Total K-mers | ABACABA Avoidmers | Expected Probability (%) |
|---|---|---|---|
| 6 | 4,096 | 4,096 | 100.00 |
| 7 | 16,384 | 16,320 | 99.609375 |
| 8 | 65,536 | 64,812 | 98.895263671875 |
| 9 | 262,144 | 256,536 | 97.8607177734375 |
| 10 | 1,048,576 | 1,011,948 | 96.50688171386719 |
| 11 | 4,194,304 | 3,978,168 | 94.84691619873047 |
| 12 | 16,777,216 | 15,586,224 | 92.9011344909668 |
| 13 | 67,108,864 | 60,859,008 | 90.6869888305664 |
| 14 | 268,435,456 | 236,831,208 | 88.22650015354156 |

**Supplementary Table 3: Longest Zimin avoidmers for k-mer lengths of 100bps or longer.**

| Chromosome | Start | End | Sequence | K-mer length |
|---|---|---|---|---|
| chr1 | 227121938 | 227122040 | gagtagacttggcgtaattcttaacagtcatggaattgttgcaatggtgaatgggcattggctttaacttaaaagttaccagctgcattagcccctatcaag | 102 |
| chr1 | 227121943 | 227122046 | gacttggcgtaattcttaacagtcatggaattgttgcaatggtgaatggcattggctttaacttaaaagttaccagctgcattagcccctatcaagagagtc | 103 |
| chr14 | 88541787 | 88541888 | aaggcaaagtcaacgttggctgaagccaggcccggtgatggttactcatgttcacataagcagcgttaggatccatggcctggacggtccagaggccgct | 101 |
| chr15 | 94141278 | 94141380 | ttcggtcaattggcaatcctgcactaggtggcaagtctatcaaaaatggattacctgggtaccatggcatttgcctgtagtcccagttactcaggaggctga | 102 |
| chr17 | 29407135 | 29407238 | gtcatggaaaaagccctggaattggaattgaatgatcctgagattatatcacagctctgcaacttactagctgtgtgacctaggtcagttggcttaacctttc | 103 |
| chr17 | 37605751 | 37605852 | cctgtaatcctagcactttggaaggccaaggtggatggataacttgaggtcaggggttcaagaccagcctggccaacatggcaaaaccctatctctactaa | 101 |
| chr19 | 8202184 | 8202285 | ctacggcaggttattcaattcatcgttgcttggatcaaaaacaaccgcctagaccaggcgtggtggctcaggtctgtaatcccagcattttgggaggctga | 101 |
| chr5 | 89164534 | 89164634 | aattcagtcgttatcaattcgggccaggtaccaaagtcactaacgtgtccaggttaactcctggaaggcagcaacattcttgcttttgatggtacgaata | 100 |
| chr7 | 57171035 | 57171139 | tcgttattgttaaggcataatgcagctggcgacttggagttgaagccattgctcatttaccattctgaaaaatcctagggcccttcagaattaagctacaccta | 104 |
| chr7 | 64843506 | 64843610 | taggtgtagcttaattctgaagggcccctaggattttcagaatggtaaatgagcaatggcttcaactccaagtcgccagctgcattatgccttaacaataacga | 104 |
| chr9 | 8763100 | 8763200 | aagctccggagttctaacgttgttagggaattaaggtccaatcaccaagcctagtccatgcttaggaatttccaaggtacttatacaggcaagtgaaagg | 100 |
| chrX | 118460314 | 118460414 | taaggattgcggtagctgccaatgccctgaggcagtggggccaacctaaagccaccaagtccagtcttttggataagtatgctaccggccgatgccaaga | 100 |
| chrX | 118526160 | 118526260 | tcttggcatcggccggtagcatacttatccaaaagactggacttggtggctttaggttggccccactgcctcagggcattggcagctaccgcaatcctta | 100 |

**Supplementary Table 4: Subsets of the most frequent Zimin avoidmers for selected compartments.** The two color-highlighted sequences in Hsat1B differ by only 1bp.

| Sequence | Length | Occurrences | Compartment |
|---|---|---|---|
| tcttcaccttgtgatccccttgccttggcctccaaatttgctgggattacaggcctgagccaagatcc**a**tatt | 73 | 4,212 | Hsat1B |
| tcttcaccttgtgatccccttgccttggcctccaaatttgctgggattacaggcctgagccaagatcc**g**tatt | 73 | 277 | Hsat1B |
| tcttcaccttgtgatccccttgccttggcctccaaatttgctgggattacaggcctgagccaagatccatatt | 73 | 2,120 | Genic |
| ggaatcgcaaggaattgatgtgaacggaacggaatggaatggaatccaaagg | 52 | 2,074 | Genic |
| ggaatcgcaaggaattgatgtgaacggaacggaatggaatggaatccaaagg | 52 | 3,567 | Hsat3 |
| aatggactcctttggaatggtgtagtatgcaatgcaatcgactggcagggaatcaaaaggaat | 55 | 1,424 | Hsat3 |
| aatggtctagtatgcaatgcaatcgactggcagggaatcaaaaggaatgtaatcg | 55 | 1,311 | Hsat3 |
| gtagtatgcaatgcaatcgactggcagggaatcaaaaggaatgtaatggaat | 52 | 780 | Hsat3 |
| aatggactcgtttggaatggtctagtatgcaatgcaatcgactggcagggaatcaaaaggaat | 63 | 713 | Hsat3 |
| aatggtctagtatgcaatgcaatcgactggcagggaatcaaaaggaatgtaatcg | 55 | 1,311 | Hsat3 |
| gattccattgggttcaattcaatgatgattacattggattccgttctatg | 50 | 1,428 | Hsat2 |
| attcgcttgctttcgatgatgattccacttgagtccgttagaagattctattcaattacattcc | 64 | 910 | Hsat2 |
| ttccattcgcttgctttcgatgatgattccacttgagtccgttagaagattctattcaattacatt | 66 | 905 | Hsat2 |
| ttccacttgagtccgttagaagattctattcaattacattccatgacgattccg | 54 | 902 | Hsat2 |
| ccgttagaagattctattcaattacattccatgacgattccgttcgagtcca | 52 | 878 | Hsat2 |
| gatcatgttgttctttcggagtaacccctacttccagaataaagtgattaccaaggaat | 59 | 43 | CDS |
| cgggctatcactggcagttcggtgtcggagaacgcggccattgccatggctggaatagccaagctctttg | 70 | 31 | CDS |
| gttgctccaatacgtaaaaggcacttctgtagggctggcatgagtcagtcagttcaagacaacctgaagga | 71 | 17 | CDS |

**Code Availability**

The GitHub code and all the related material for this work is provided at: https://github.com/Georgakopoulos-Soares-lab/Avoidmers-DNA/

**Funding**

Research reported in this publication was supported by the National Institute of General Medical Sciences of the National Institutes of Health under award number R35GM155468.